\documentclass[preprint, aps, prb, onecolumn, superscriptaddress, floatfix, altaffilletter]{revtex4}

\usepackage{graphicx}
\usepackage{dcolumn}
\usepackage{bm}
\usepackage{amsmath}
\usepackage{multirow}
\usepackage{hhline}

\begin{document}

\title{Computational generation of voids in $a$-Si and $a$-Si:H by cavitation at low density}

\author{Enrique Guerrero} \email{eguerrero23@ucmerced.edu}
\affiliation{Department of Physics, University of California, Merced, Merced, CA 95343}
\author{David A. Strubbe} \email{dstrubbe@ucmerced.edu}
\affiliation{Department of Physics, University of California, Merced, Merced, CA 95343}

\begin{abstract}


  Use of amorphous silicon ($a$-Si) and hydrogenated amorphous silicon ($a$-Si:H) in photovoltaics has been limited by light-induced degradation (the Staebler-Wronski effect) and low hole mobilities, and voids have been implicated in both problems. Accurately modeling the void microstructure is critical to theoretically understanding the cause of these issues. Previous methods of modeling voids have involved removing atoms according to an {\it a priori} idea of void structure and/or using computationally expensive molecular dynamics. We propose a new fast and unbiased approach based on the established and efficient Wooten-Winer-Weaire (WWW) Monte Carlo method, by using a range of fixed densities to generate equilibrium structures of $a$-Si and $a$-Si:H that maintain 4-coordination. We find a smooth evolution in bond lengths, bond angles, and bond angle deviations $\Delta \theta$ as the density is changed around the equilibrium value of $4.9\times10^{22}\ $atoms/cm$^3$. However, a significant change occurs at densities below $4.3\times10^{22}\ $atoms/cm$^3$, where voids begin to form to relieve tensile stress, akin to a cavitation process in liquids. We find both small voids (radius $\sim$3~\AA) and larger ones (up to 7 \AA), which compare well with available experimental data. The voids have an influence on atomic structure up to 4 \AA\ beyond the void surface and are associated with decreasing structural order, measured by $\Delta\theta$. We also observe an increasing medium-range dihedral order with increasing density. Our method allows fast generation of statistical ensembles, resembles a physical process during experimental deposition, and provides a set of void structures for further studies of their effects on degradation, hole mobility, two-level systems, thermal transport, and elastic properties. The basic concept of generating voids at low density is applicable to other amorphous materials.

\end{abstract}

\date{\today}
\maketitle

\section{Introduction}
Amorphous silicon ($a$-Si) is a cheap and flexible semiconductor\cite{Street} used in ultra-reflective mirrors,\cite{Steinlechner} thin-film transistors,\cite{Nathan} and solar cells.\cite{Shah} A resurgence in interest in the material comes from the hetero-junction with intrinsic thin-layer (HIT) cell, a $c$-Si/$a$-Si tandem solar cell with high efficiency comparable to traditional crystalline silicon ($c$-Si) solar cells.\cite{Taguchi,Bush} Unfortunately, fielded HIT cells suffer from twice the degradation rate of single-crystal Si cells.\cite{Jordan,Ishii} This increased rate is likely due to the light-induced Staebler-Wronski degradation\cite{SWE} of $a$-Si:H, which has been attributed to the breaking of Si-H bonds at small voids.\cite{Fehr} Voids play an important role in other properties of $a$-Si as well. Low hole mobility is a key shortcoming in $a$-Si and it has been linked to tensile stress and hence voids in the material.\cite{Johlin2014} Density and H-content (both associated with voids) significantly affect elastic properties of $a$-Si.\cite{Jiang} Thermal conductivity is an important issue for thermoelectrics (where it should be low) and for optoelectronic devices (where it should be high), and experiments show important effects of microstructure on thermal conductivity of $a$-Si.\cite{Jugdersuren} Simulations have predicted that porosity on the order of 1 nm significantly lowers the thermal conductivity.\cite{Park} A better understanding of void structures and properties could open up new strategies for engineering thermal transport in $a$-Si with porosity, as has been explored in $c$-Si.\cite{Romano} A final area of application is two-level systems, believed to be responsible for low-temperature contributions to the specific heat. Voids may cause two-level systems, \cite{Queen} or provide vibrational modes that mimic the effects of two-level systems.\cite{Nakhmanson} Due to the omnipresence of voids in amorphous systems, further study of the properties of voids in $a$-Si and $a$-Si:H is crucial to understanding the macroscopic behavior of these materials. $a$-Si and $a$-Si:H are some of the most studied amorphous materials, and can serve as model systems for testing ideas about amorphous materials in general.

Void content, along with density and intrinsic stress, is dependent on deposition conditions\cite{Johlin2012,Vanecek} but is found in essentially all $a$-Si samples. The existence of microvoids has been observed using small-angle electron scattering\cite{Moss} and small-angle x-ray scattering.\cite{Mahan_1989} These methods find density-deficient regions in $a$-Si:H and attribute them to voids. Nuclear magnetic resonance\cite{Baum} and infrared (IR) absorption\cite{Ouwens,Smets} techniques suggest that H atoms tend to cluster, perhaps even in the form of molecular H$_2$.\cite{Chabal} Increasing H concentrations can increase the amount of H clustering\cite{Smets} and decrease the Young's modulus.\cite{Jiang} H effusion has been used to indirectly measure voids, but may not distinguish between microvoids or interconnected low-density regions; instead He implantation and temperature-mediated effusion can study voids with divacancy-level resolution.\cite{Beyer} Experimental void research in non-hydrogenated $a$-Si is sparse, but a-Si may in fact contain fewer voids than a-Si:H.\cite{Williamson,Remes} Recent experiments on density variation and voids in $a$-Si have attempted to shed light on the origins of two-level systems.\cite{Jacks,Molina,Molina_2018}
This zoo of experimental measurements yields void sizes to be anywhere from the size of divacancies,\cite{Beyer} about 3 \AA\ in radius, to as large as 20 nm.\cite{Vanecek} Void number density ranges from $10^{18}-10^{20}$ cm$^{-3}$.\cite{Beyer,Mahan} Experimental conclusions can differ widely: IR measurements suggest that divacancies dominate the void content\cite{Smets} while He effusion shows larger voids are more prevalent.\cite{Beyer} Different deposition conditions of course can give different microstructures. 

The common method for void generation in computational $a$-Si:H is atomic removal \cite{Chakraborty,Pedersen,Nakhmanson,Kim,Biswas_2007,Paudel}: $a$-Si coordinates are generated, a choice of Si atoms are removed, and dangling bonds may be passivated by H-insertion. This method has been used to model void effects on the vibrational specific heat,\cite{Nakhmanson} small-angle x-ray scattering,\cite{Paudel} hydrogen evolution,\cite{Chakraborty} and paracrystalline structures.\cite{Biswas_2007} Si removal can generate voids of controllable shape and size, but it has an inherent bias: the surrounding structure has a limited ability to reconstruct, and there will be dangling bonds left behind (4 by a monovacancy, 6 for a divacancy, etc.). Small voids are known to be stable in $a$-Si and found in the equilibrium structure. Pedersen {\it et al.}\cite{Pedersen} have used this idea to generate realistic, low-energy $a$-Si structures by a grand-canonical Monte Carlo method in which atoms can be removed to find lowest-energy densities and bond topology. Biswas {\it et al.}\cite{Biswas} used a metadynamics approach\cite{Biswas_2016} as an alternative to atomic removal, finding voids as a product of biasing structures to fit bonding defect constraints. This method has been used to study microvoids of radii 6-12 \AA\ in large cells (7000 Si atoms), carefully constructing a description of the complex-shaped void network.\cite{BiswasVoids} Such a calculation relies on assumed associations between voids and coordination defects. There have been relatively few studies so far of defects in non-hydrogenated $a$-Si, \textit{i.e.} of non-passivated voids.\cite{Kim,Nakhmanson,Biswas_2007,Paudel}

We take a complementary approach to the previous work, using only the WWW method. We explore voids with radii up to 7 \AA\ by performing Monte Carlo simulations at a fixed low density, and allowing them to form in the approach to equilibrium. While this technique is a usage of the well-established WWW method, it does not seem to have been explored before. Our approach of annealing at constant volume and number of atoms potentially is more closely connected to the physical processes of chemical vapor deposition growth,\cite{Johlin2014} in which initially deposited Si (and H) atoms on a surface at elevated temperature undergo an annealing process to form the final structure.\cite{Smets} The melt-quench approach \cite{Kluge} could potentially be used to prepare voids, but voids may be controlled more by bubble formation in the liquid than the properties of the solid network. Our method avoids expensive density functional theory (DFT) or melt-quench molecular dynamics during structure generation, and is computationally simple and efficient. The general idea of producing voids by generating structures at low density does not rely on any particular feature of $a$-Si or $a$-Si:H, and therefore is generalizable to other amorphous materials.

Many works choose either $a$-Si or $a$-Si:H as the material of interest; we have studied voids present in both materials due to the transferability of our methods. Our aim is to generate structures with voids for use in studying the effects on light-induced degradation\cite{Zimanyi} and other optoelectronic properties of $a$-Si:H. We use the Wooten-Winer-Weaire\cite{WWW} method to generate ensembles of $a$-Si and $a$-Si:H at 10\% hydrogen content, as is commonly used for electronic devices.\cite{Street} We modify the WWW algorithm and observe the formation of voids in the equilibrium structures at a given density, rather than explicitly removing atoms. Typical $a$-Si simulations only consider experimental densities; we instead systematically vary our density and find that the stochastic evolution of our structures favors void formation at low densities. We consider structures with densities as low as $3.4\times 10^{22}$ at/cm$^3$, well below device-quality $a$-Si:H, because they each represent a void and the region around it, which may be embedded in a matrix of higher density. The structures around 4.8-$5.0\times 10^{22}$ at/cm$^3$ model regions of material with the typical density of the continuous random network.

The paper is organized as follows. In Section II, we describe our methods. In section IIA, we describe the modifications to the WWW algorithm used to generate structures as well as overcoming difficulties produced by varying density. Section IIB covers the details of DFT calculations using Quantum ESPRESSO.\cite{QE-2009} Section IIC discusses how to characterize voids with a Voronoi tesselation and how we correlate those voids to structural effects. Section III show results on overall structural changes and then the localized changes near to voids. In Section IV, we conclude.

\section{Methods}
\subsection{CHASSM}
We use the Computationally Hydrogenated Amorphous Semiconductor Structure Maker (CHASSM)\cite{CHASSM} code, which implements the WWW Monte Carlo approach \cite{WWW}. The ordinary WWW process is described as follows: (1.) Create a periodic $c$-Si structure. (2.) Propose a bond-switch between neighboring bonds and relax the new structure's atomic coordinates. (3.) Compare the proposed structure's energy to the previous structure using the Boltzmann factor $e^{-\Delta E/k_B T}$ to decide the probability of accepting such a move. (4.) Return to step (2.). CHASSM makes two changes to the initial crystal: we triaxially strain the initial crystal to a target density; and we delete random Si-Si bonds to create a pair of Si-H bonds,\cite{Wagner} up to a desired number of H atoms in the sample. This approach avoids any {\it a priori} ideas of where H atoms should go, as involved in schemes of identifying and passivating dangling bonds. \cite{Biswas} Our Keating potential does not have any terms involving the H atoms. In the final structure, an H atom bonded to a given Si atom is considered to be located in a position opposite the Si atoms bonded to it. Structures of $a$-Si:H from this code, generated in the usual way with a fixed density, have been used to study barriers to bond-switching in the Staebler-Wronski effect,\cite{Wagner} strain-induced shifts in Raman peaks,\cite{Strubbe} optical absorption,\cite{Raghunathan} and nanocrystalline sites in a-Si,\cite{Mueller} validated with a variety of properties. Note that given the significant energetic and entropic barriers between different amorphous structures, straining structures to a different density and simply relaxing (as for studying effects of small strain \cite{Strubbe}) would not produce as much structural variation as we find here, and would not correspond to the experimental growth to different densities which we are targeting.

We use the Keating classical potential\cite{Keating} as the energy in the Boltzmann factor. It relies on a predetermined bond table, not a set of distance-based nearest neighbors, to decide which atoms interact. The Keating potential is as follows\cite{Keating}:
\begin{equation}
U=\frac{3\alpha}{16\delta^2}\sum_{i}^{N_{atoms}}\sum_{j}^{N_{b, i}}\bigg(\Big(\lvert \boldsymbol{r}_{ij}\rvert^2 -\delta^2\Big)^2 + \frac{2\beta}{\alpha}\sum_{k>j}^{N_b, i}\Big(\boldsymbol{r}_{ij}\cdot \boldsymbol{r}_{ik} + \frac{\delta^2}{3}\Big)^2\bigg)
\end{equation}
%
%
%
where $\alpha$ and $\beta$ are bond length and angle force constants, $\delta$ is the equilibrium Si-Si bond length, $N_{b,i}$ is the number of bonds to atom $i$ (fixed at 4 for $a$-Si), and $\boldsymbol{r}_{ij}$ is the bond vector from atom $i$ to its $j$th-bonded atom. We have set $\alpha = 2.965$ eV/\AA$^2$, $\beta = 0.845$ eV/\AA$^2$, and $\delta = 2.35$ \AA, to match experimental values for $c$-Si as used by Barkema and Mousseau.\cite{Barkema} It is interesting to note that the bond-angle term is essentially the tetrahedral order parameter used in systems such as amorphous ice.\cite{Tetrahedral}

While the Keating potential is a fairly crude potential, and more accurate Si potentials have been developed such as Tersoff \cite{Tersoff} and Stillinger-Weber,\cite{SWPotential} the Keating potential does have a crucial feature -- it is based on a concept of a bonding network, which is needed to define the Monte Carlo move as a bond switch. The Tersoff and Stillinger-Weber potentials are distance-based without a definition of bonds, and therefore cannot replace Keating in the WWW process (though additional distance-based terms can be added \cite{vonAlfthan}). The deviations between Keating and the other potentials are significant mainly for structures far from equilibrium, such as during the amorphization process, but the accuracy of these intermediate steps is not important for our final structures. Small deficiencies in the final structures are corrected by the DFT relaxation.

We allow the structure to evolve under a changing temperature ($T$) to ensure escape from the crystal phase and local minimization in the amorphous regime of the energy landscape. The temperature profile consists of three phases. An initial `randomization' phase of 800 switch attempts/atom at high $T$ (about 0.8 eV) is used to escape the crystal barrier while highly distorting the bonding network. The next `anneal' phase consists of 100 switch attempts/atom at decreasing $T$ (0.8 to 0.4 eV in intervals of 0.002$-$0.05 eV); this slow cooling allows improvement of the bonding network while the system traverses small barriers in the rough landscape\cite{Charbonneau,Stillinger} to reach local minima. Finally, we `quench' (100 switch attempts/atom at $T=0$) to relax and ensure the system is at a local minimum.

\begin{figure}
	\includegraphics[width=250px]{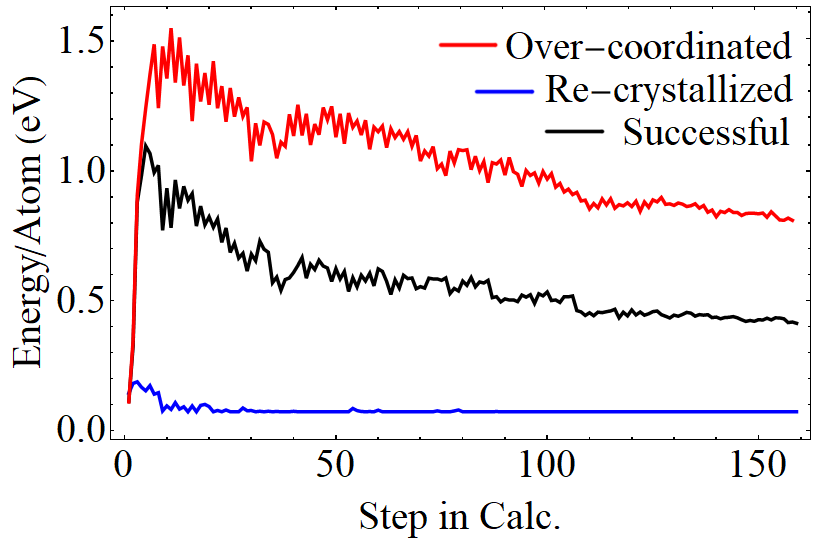}
	\caption{Keating energy throughout a CHASSM calculation at a density of $4.5\times10^{22}\ $at/cm$^3$. The first 8-10 steps are high-$T$ randomization. If the structure fails to obtain enough energy to escape the barrier to the amorphous phase, it re-crystallizes to a strained $c$-Si (blue). If the structure randomizes at too high a $T$, it does not relax to a reasonable energy (red) or bonding network. A run producing a desired realistic amorphous structure has an intermediate behavior (black).}
	\label{fig:EnStep}
\end{figure}

If the randomization $T$ is too low, the network will not be sufficiently perturbed from a perfect lattice and reverts to a crystal,\cite{WWW} dropping to a low energy as shown in Fig. \ref{fig:EnStep}. We encountered an opposing problem: if the randomization $T$ is too high, the bonding network distorts too far from a physical one to be annealed. Since the Keating potential does not rely on nearest neighbors, atoms may be within coordination shells but have no interaction if they are not ``bonded" according to the bond table.\cite{vonAlfthan} Structures with too large a randomization temperature may be artificially over-coordinated: they may have 9+ atoms within the first coordination shell but only four Keating bonds. Structures of this kind will have very high energies shown in Fig. \ref{fig:EnStep}. To remedy this, we find ideal randomization $T$ empirically: we randomize structures at variable temperatures for 1000 steps for each density, and the smallest temperatures that escape the crystal phase are chosen. Ideal temperatures minimize the number of failed structures due to re-crystallization or artificial coordination. Increasing the density increases barriers and requires a higher initial temperature. We find the ideal temperature to be $T=0.82\ $eV$ - (\rho-\rho_0)\ 0.18\ $eV/(10$^{22}\ $at/cm$^3$), where $\rho-\rho_0\ $is the difference between the density $\rho$ and the relaxed crystal density, $\rho_0=5.0\times$10$^{22}$ at/cm$^3$. At densities below $3.4\times10^{22}\ $at/cm$^3$, the $T$ required to overcome the initial barrier will always over-distort the bonding network. Densities above $5.8\times10^{22}\ $at/cm$^3\ $will always be over-coordinated; we discard structures with any atoms with 5+ atoms within the first coordination shell, since the Keating potential does not describe them well. Our densities are thus limited by those two values. Hydrogenated structures have a larger range of usable $T$ than pure a-Si structures but follow the same ideal $T$ trend, which we attribute to the more flexible bonding network when Si-Si bonds are replaced with Si-H bonds.

The WWW algorithm can be disrupted by identical bonding: after bond-switching, two atoms may end up bonded to the same set of four atoms but not to each other. These atoms will inevitably relax to the same location yet feel no mutual interaction, which is strongly unphysical. The likelihood of such events increases with the system size and is particularly important to address for structures of 1000+ atoms. We solved this by rejecting switch attempts that would cause two atoms to have the same set of bonds. It could also be remedied by including distance-based repulsive terms to the potential.\cite{vonAlfthan,SWPotential}

We use CHASSM to generate ensembles of structures at variable densities of both $a$-Si (Si$_{216}$) and $a$-Si:H (Si$_{216}$H$_{20}$) from 3.4 to $5.6\times10^{22}\ $at/cm$^3$ in intervals of $0.16\times10^{22}\ $ at/cm$^3$. 10 structures per density are sampled to be further relaxed using plane-wave DFT. Stresses of $\pm$1 GPa are common in $a$-Si:H, \cite{Johlin2014} and in this work we reach 5 GPa. Cells are fixed as cubic, with lattice constant ranging from 15.6 \AA\ to 18.5 \AA\ at the highest and lowest densities respectively. Structural parameters of DFT-relaxed structures are calculated, and the error bars displayed are the standard errors of the population of 10 structures. The structural parameters of the original CHASSM structures in the full data set (bond lengths, bond angles, bond angle deviations) are found to be very similar to the results of DFT relaxation, and are not shown. Pressure results are from a stress calculation in CHASSM implemented in the approach for classical potentials in periodic systems detailed in Ref. \onlinecite{Plimpton}, using their equations (28) and (29).

\subsection{DFT}
We use Quantum ESPRESSO\cite{QE-2009} to perform fixed-cell relaxations at the $\Gamma$ point using the PBE exchange-correlation potential\cite{PBE} and ultrasoft pseudopotentials (USPP).\cite{USPP,PPURLs} We set the wavefunction kinetic energy cutoff to 38 Ry and 46 Ry for $a$-Si and $a$-Si:H respectively. Charge density cutoffs (requiring special care for USPP) were set to 151 Ry and 221 Ry for $a$-Si and $a$-Si:H. CHASSM structures were relaxed until forces and energies were converged to $10^{-4}\ $Ry/Bohr$^2\ $and $10^{-4}\ $Ry respectively. These values were chosen because lowering thresholds only affected the atomic positions by less than $10^{-6}\ $\AA. Structures at very low and high densities required smearing to converge the self-consistent cycle, possibly due to unpaired electrons at floating or dangling bonds. For calculations of the relaxed density, we perform variable-cell relaxations until the stress tensor elements are below $\pm 0.01\ $kbar. $a$-Si structures below $3.6\times10^{22}\ $at/cm$^3$ did not reliably converge self-consistent field cycles. After DFT relaxation, we consider atoms within 2.8 \AA\ of each other to be bonded; bond lengths, angles, and dihedrals are computed from this bonding network.  Our full set of CHASSM and DFT structures for each density, $a$-Si and $a$-Si:H, are provided in the supplementary material.\cite{SM}

\subsection{Void Characterization}
We delegate our void characterization to Zeo++,\cite{Zeo} an open-source code developed to study the structure of void channels in zeolites. The code's pore-size distribution\cite{ZeoPSD} function samples ``test points" in the material and records the radius of the largest sphere encapsulating each point without touching any atoms. Note that this method interprets what could be considered a complex-shaped void (as in Ref. \onlinecite{Biswas}) as several spherical voids. We consider our characterization to be appropriate if we are not concerned with details of the voids' surface structure.

\begin{figure}
	\includegraphics[width=180px]{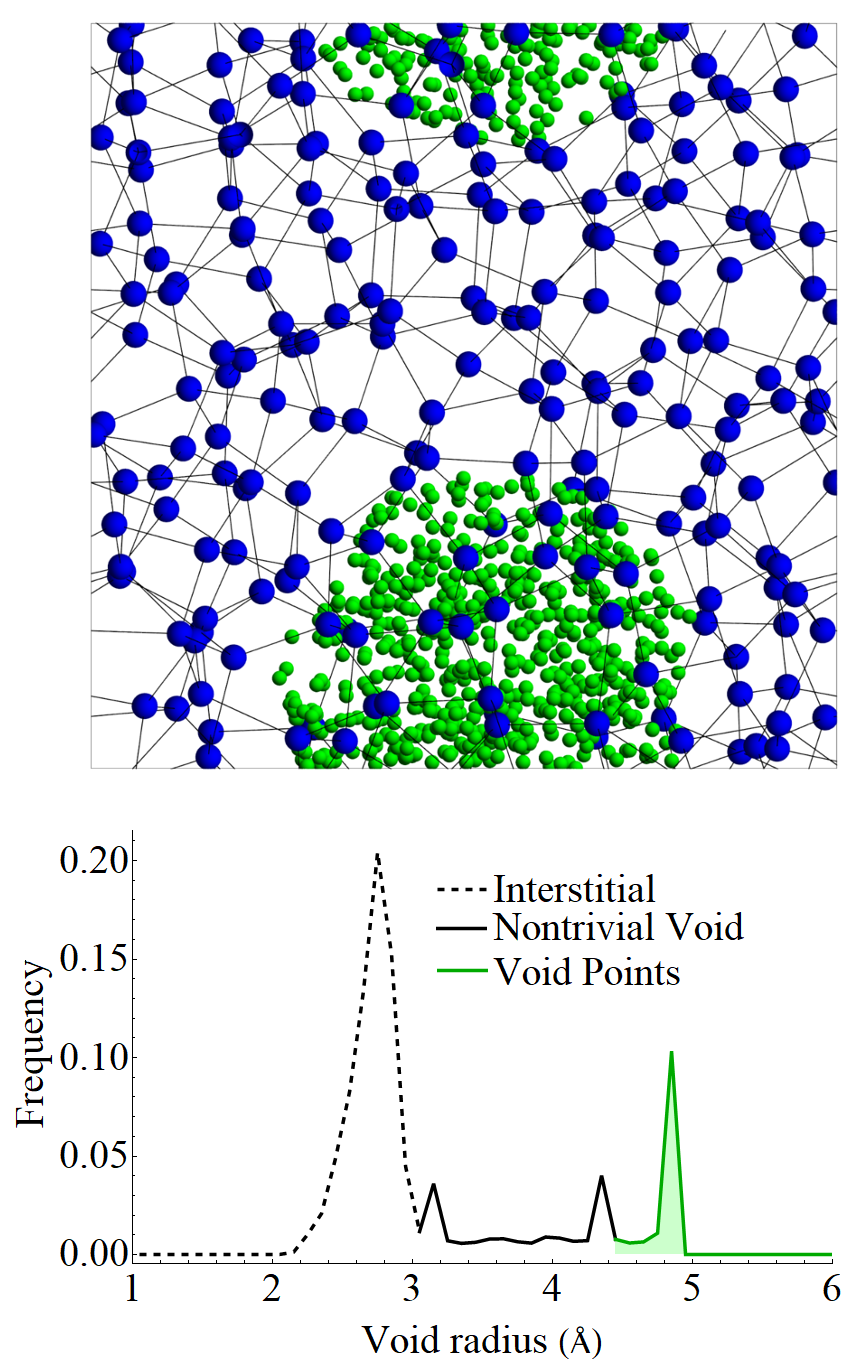}
	\caption{(top) An example low-density ($4.3\times10^{22}\ $at/cm$^3$) $a$-Si structure with a large void. The green void points fill in the largest 10\% of the void size distribution. (bottom) The pore size histogram of a low-density ($4.05\times10^{22}\ $at/cm$^3$) post-DFT structure. Large voids (4.9 \AA) and interstitial voids (2.5 \AA, dashed line) appear as strong signals in this histogram. The area underneath the solid region constitutes the void volume, excluding interstitials. Only void points belonging to the largest 10\% of voids (green) are considered for the void proximity ($r_v$) analysis.}
	\label{fig:PSD}
\end{figure}

We have set the atomic radii and probe size to zero in Zeo++, and we have only considered Si atoms for void analysis to be able to directly compare $a$-Si to $a$-Si:H. All structures show a strong peak of interstitial-like voids (Fig. \ref{fig:PSD}), a broadened version of the single crystal peak which appears at 2.4 \AA. Low-density voids will appear as one or more peaks beyond the interstitial peak. To quantify the total void volume, we ignore the interstitial peak from the distribution. At the lowest densities we attain up to 30\% void volume, showing the efficiency of this method for generating ensembles of voids and their neighborhoods. By contrast the study of Paudel {\it et al.} produced structures of only $\sim 0.3$\% void volume.\cite{Paudel} Void concentration in our lowest-density calculations is two orders of magnitude larger than that found around the equilibrium density by Biswas {\it et al.}\cite{Biswas} The void sizes in our calculation are necessarily limited to be smaller than the supercell we have used; larger supercells would allow larger voids.

To find the renormalized densities of the non-void parts of the structure, we set the radii of Si atoms to 2.21 \AA, slightly above the Van der Waals radius. Values much larger than this would leave no interstitial volume outside the spheres; values much smaller would lead to the whole structure outside the atoms being taken as a single connected void. Renormalized densities are calculated as $\rho_{norm}=\rho/N_{atomic}$, where $N_{atomic}$ is the proportion of test points that fall within 2.21 \AA\ of any Si atoms. Mono- and divacancies have been studied extensively by IR spectroscopy, \cite{Smets} helium effusion, \cite{Beyer} and computational studies.\cite{Kim, Chakraborty} If we generate mono- and divacancies in $c$-Si, our approach gives void radii of 2.5 and 3.0 \AA\ respectively. However, these values fall within the range for interstitial voids present in all samples, and thus we do not see mono- and divacancies as distinguishable structures. For densities above $4.0\times10^{22}\ $at/cm$^3$, voids containing a single H atom are similarly indistinguishable from the interstitial and monovacancy size distributions. Some previous work has raised the question of the ``cavity'' around the H atom in $a$-Si:H\cite{Beyer} -- our calculation shows that a single H atom fits into even a dense Si-Si bonding network without causing any significant distortion.

We locate large voids by considering test points corresponding to the largest 10\% of spheres in a given structure to be that structure’s ``void points" (pictured in Fig. \ref{fig:PSD}). We assign a void proximity measure to every atom, $r_v$, defined as the shortest distance from that atom's center to a void point. We associate this distance with structural parameters to study how far the void's influence extends into the material.

\section{Results and Discussion}

We can probe the differences between pre- and post-DFT bond topologies to assess the validity of CHASSM structures. Any atoms whose local bonding has been readjusted (whether by a broken or new bond) is counted as a bond correction. Away from relaxed densities, these events are more common, at worst 3\% compared to the 0\% near relaxed densities. Atomic positions are corrected by DFT on average by 0.04 \AA. We take these as evidence that DFT preserves the topology created by the Keating potential reasonably well, except at the most extreme densities we have studied.

\begin{table*}[t]
\begin{tabular} {| l || r | r | r || r | r | r || r | r | r |}
\multicolumn{1}{c}{} & 
\multicolumn{3}{c}{\bfseries CHASSM} & 
\multicolumn{3}{c}{\bfseries CHASSM} & 
\multicolumn{3}{c}{\bfseries Exp't} \\
\multicolumn{1}{c}{} & 
\multicolumn{3}{c}{} & 
\multicolumn{3}{c}{\bfseries +DFT	} & 
\multicolumn{3}{c}{} \\
\hline
\hline
 & 
{$c$-Si} & 
{$a$-Si} & 
{$a$-Si:H} & 
{$c$-Si} & 
{$a$-Si} & 
{$a$-Si:H} & 
{$c$-Si} & 
{$a$-Si} & 
{$a$-Si:H} \\
\hline
$\rho_0$, $10^{22}\ $at/cm$^3$ & 
5.01 & 5.12 & 5.07 & 
4.87 & 4.78 & 4.67 & 
5.01\cite{Hopcroft} & 4.9\cite{Witvrouw} & 4.9\cite{Kuschnereit}\\
$Y$, GPa & 
162 & 180 & 166 & 
153 & 138 & 129 & 
165\cite{Hopcroft} & 140\cite{Witvrouw} & 126\cite{Kuschnereit}\\
$B$, GPa &
97 & 77 & 64 & 
82 & 59 & 60 &
98\cite{Hopcroft} & 140\cite{Queen} & 59\cite{Tanaka}\\
$\langle r \rangle$, \AA & 
2.35 & 2.33 & 2.34 & 
2.37 & 2.36 & 2.38 & 
2.35\cite{Hopcroft} & 2.38\cite{Fortner} & 2.36\cite{Schulke}\\
$\langle \theta \rangle$, degrees & 
109.5 & 109.3 & 109.3 & 
109.5 & 109.2 & 109.1 & 
109.5 & 108.5\cite{Shao} & 108.4\cite{Fortner}\\
$\Delta \theta$, degrees & 
0 & 9.6 & 9.9 & 
0 & 10.3 & 11.0 & 
0 & 8-11\cite{Roura,Fortner} & 8-11\cite{Roura,Roorda}\\
\hline
\end{tabular}
\label{table:density}
\caption{Relaxed density $\rho_0$ and corresponding properties from CHASSM, CHASSM+DFT, and experiment, for $c$-Si, $a$-Si, and $a$-Si:H: Young's modulus ($Y$), bulk moduli ($B$), mean bond length ($\langle r \rangle$), mean bond angle ($\langle \theta \rangle$), and bond angle deviation ($\Delta \theta$). }
\end{table*}

\begin{figure}
	\includegraphics[width=250px]{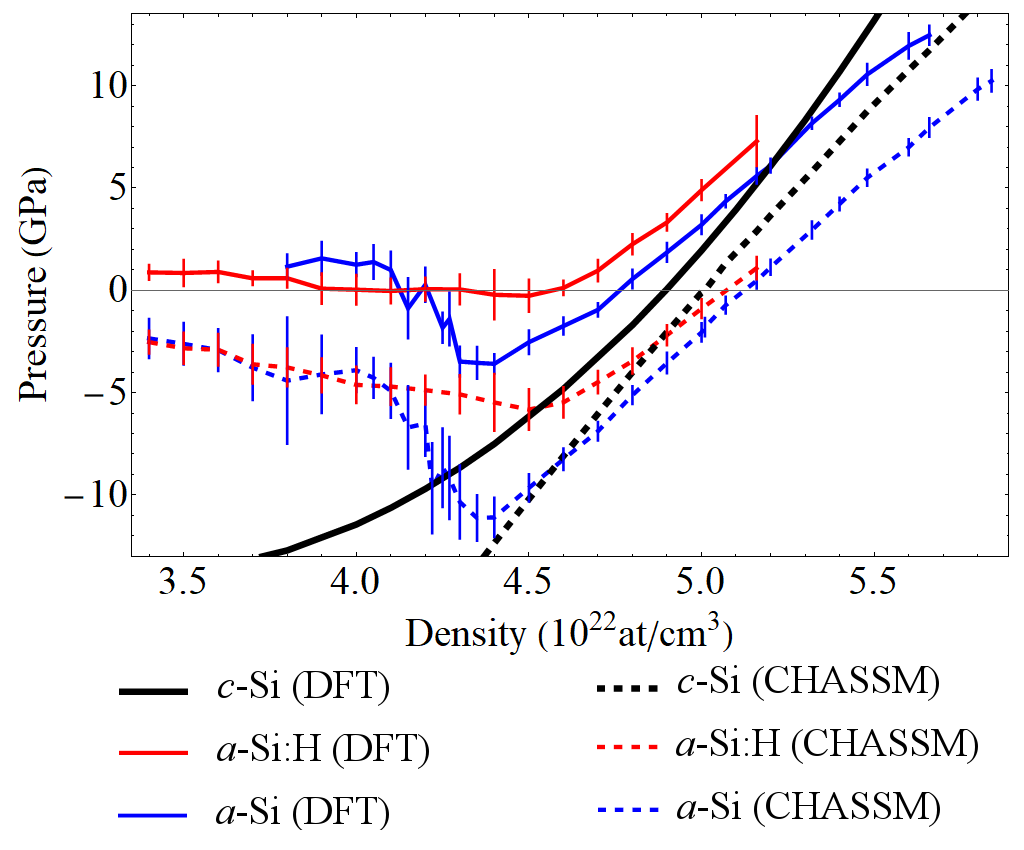}
	\caption{CHASSM (solid) and CHASSM + DFT (dashed) calculated pressures vs. densities. As density is decreased in $a$-Si and $a$-Si:H, negative pressure is induced, but then relieved near the void onset density of 4.3--$4.5\times10^{22}$ at/cm$^3$, similar to the cavitation process of bubble formation. CHASSM pressures are systematically too low compared to DFT, but have the correct trend. $a$-Si has a more abrupt transition than $a$-Si:H.}
	\label{fig:Press}
\end{figure}

We benchmark the density, elastic properties, and structural parameters at the relaxed density in Table I. The densities of both $c$-Si and $a$-Si are underestimated by PBE by $0.1\times10^{22}\ $at/cm$^3$. The relaxed $c$-Si CHASSM density (by choice of the Keating parameters $\alpha$, $\beta$, and $\delta$) matches experiment, but $a$-Si is incorrectly denser than $c$-Si, as noted in the original WWW work.\cite{WWW} This does not affect results for a fixed density. Elastic constants are described well by CHASSM only for $c$-Si near relaxed densities (Fig. \ref{fig:Press}), due to the lack of any dependence beyond harmonic in the Keating potential, but the DFT elastic constants agree well with experiment. Structural parameters agree well with experiment, and we find similar levels of agreement for $a$-Si and $a$-Si:H. All comparisons with experiment must of course take into account the substantial variation possible due to different fabrication conditions for these materials.

\begin{figure}
	\includegraphics[width=250px]{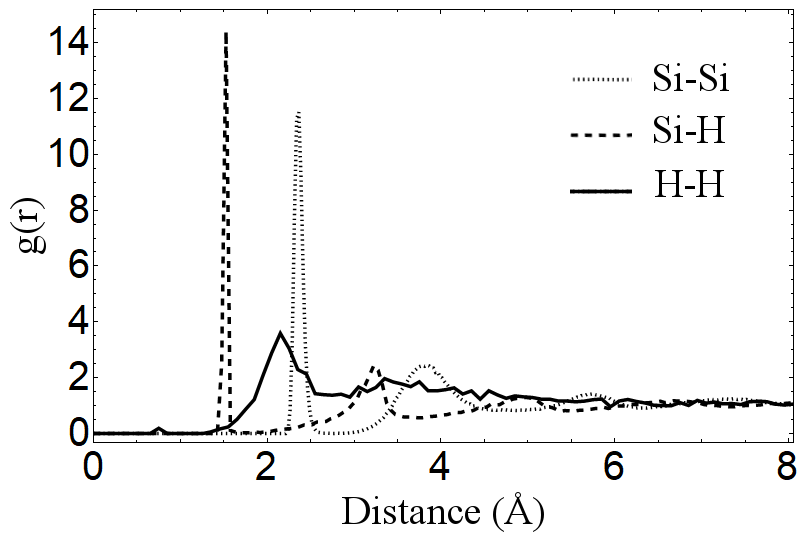}
	\caption{Averaged partial pair distribution functions, $g(r)$, for $a$-Si:H at all densities. The Si-Si $g(r)$ for $a$-Si is indistinguishable from that of $a$-Si:H. Decreasing density increases the height of the H-H 2.2 \AA\ peak, but has little effect on the other curves.}
	\label{fig:gr}
\end{figure}

Our calculated pair distributions $g(r)$ are shown in Fig. \ref{fig:gr}. We find they have little dependence on density, and the Si-Si $g(r)$ is very similar for $a$-Si and $a$-Si:H. A 2.2 \AA\ peak in the H-H pair distribution function is consistent with SiH$_2$ bonding networks found in divacancies created with molecular dynamics,\cite{Chakraborty} and with neutron scattering.\cite{Bellisent} This peak is a sign that H atoms preferentially cluster near the interior of voids.

\begin{figure}
	\includegraphics[width=200px]{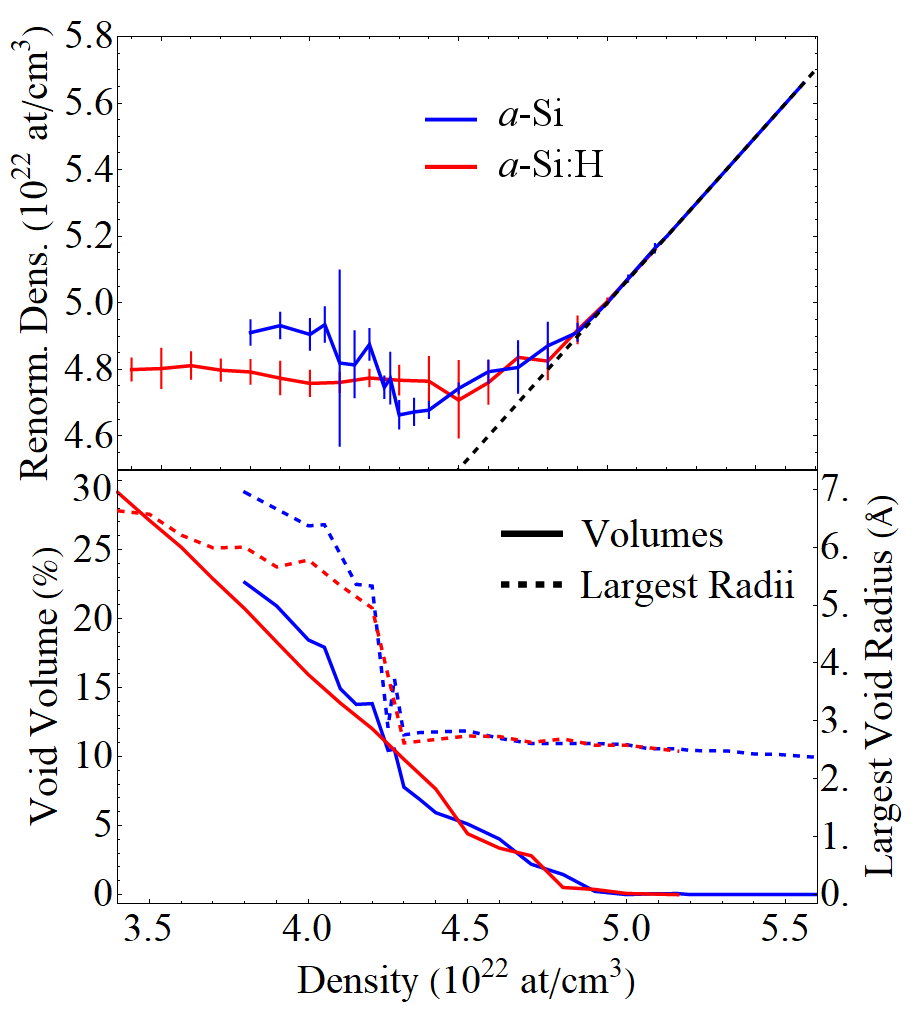}
	\caption{Renormalized densities (top) indicate the density of the non-void regions, showing that void formation allows the rest of the sample to retain a constant density. (bottom) Voids start forming at $4.3$--$4.5\times10^{22}\ $at/cm$^3$. Above the critical density, the largest void radii are about the size of the interstitial, and the total void volume is essentially zero.}
	\label{fig:voids}
\end{figure}

Pressures (minus one-third of the trace of the stress tensor) calculated using CHASSM are significantly more negative than those obtained from DFT, but they have a similar trend with a constant offset in Fig. \ref{fig:Press}. Pressures vary linearly with density above $4.5\times10^{22}\ $at/cm$^3$. A sudden drop in absolute pressure occurs at the critical density between 4.3 and $4.5\times10^{22}$ at/cm$^3$, showing stress relief. These densities are consistent with the onset of voids in Fig. \ref{fig:voids}. This behavior shows the same physical mechanism as cavitation and bubble formation at low pressures in liquids.\cite{Blander} A sharp drop in pressure at low densities is similarly observed in classical molecular dynamics simulations of water.\cite{Wang} At low densities, small voids are non-existent because they have instead coalesced into one large void or even a channel. Once voids approach the size of our supercell, they are likely to meet their periodic neighbors and form connected channels. This process is not only an artefact but could also be related to the observation of cylindrical rather than spherical cavitation in water at low enough densities.\cite{Wang} We conclude that voids have been created to relieve the global pressure caused by a highly strained bonding network. The pressure stabilizes to a constant value at the lowest densities for all data sets.

\begin{figure}
	\includegraphics[width=380px]{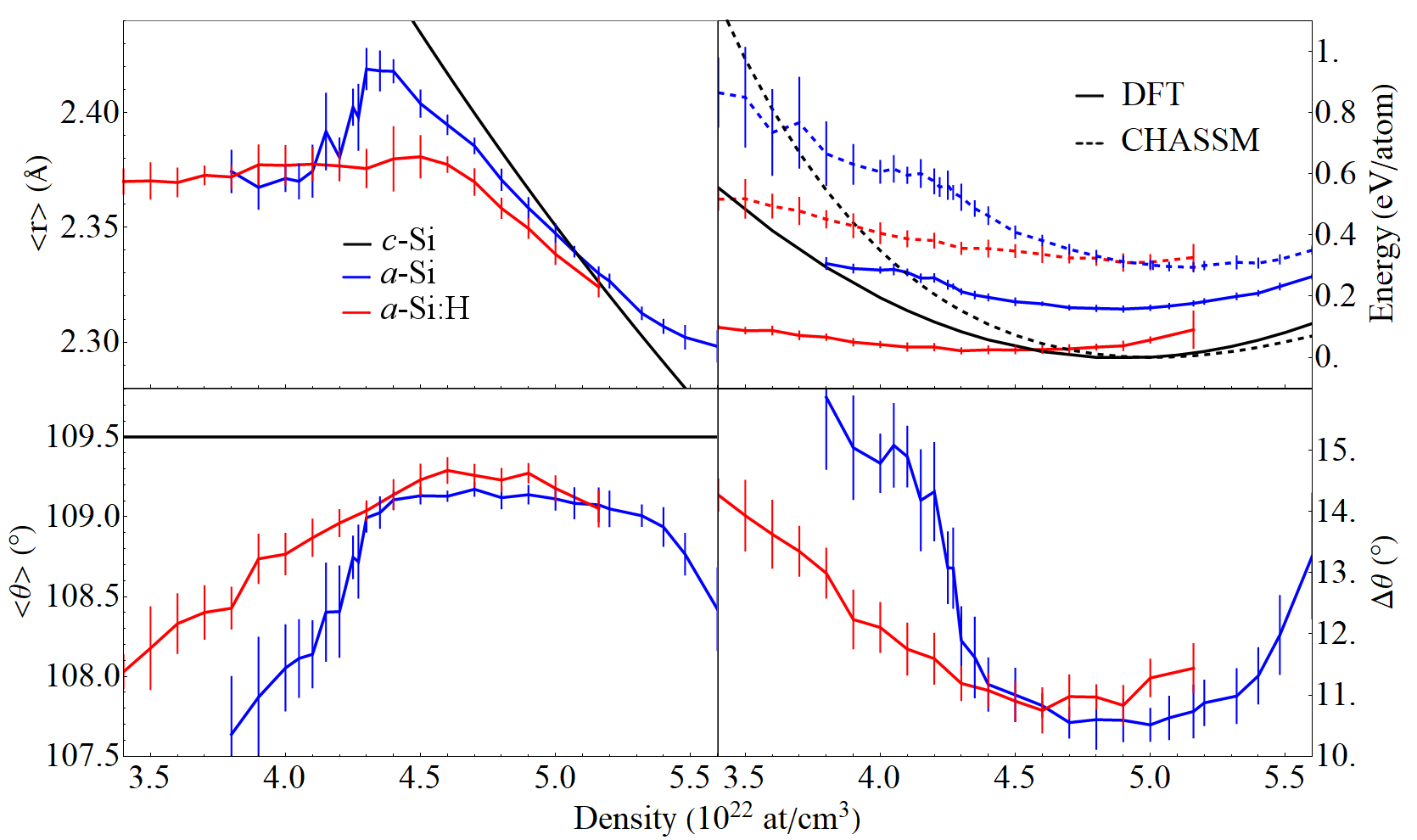}
	\caption{Evolution of structural parameters with density. Bond lengths and angles change trends around $4.3\times10^{22}\ $at/cm$^3$, the density of void onset shown in Fig. \ref{fig:voids}. Relaxed $c$-Si has a CHASSM energy of 0 eV and $\Delta \theta = 0^\circ$. $a$-Si DFT energies are relative to $c$-Si and $a$-Si:H energies are relative to the lowest $a$-Si:H energy in our data set.}
	\label{fig:Global}
\end{figure}

This picture of cavitation is reinforced by examination of the bond lengths and angles (Fig. \ref{fig:Global}), which have a transition around the critical density $4.3\times10^{22}\ $at/cm$^3$. Bond lengths in $a$-Si increase as density is decreased, but then decrease again back to the relaxed value after stress relief with void formation. The small magnitude of bond length changes are consistent with the results of Jacks and Molina-Ruiz {\it et al.},\cite{Jacks,Molina} from electron-energy loss spectroscopy (EELS). Overall, $a$-Si:H structures react more smoothly to strain because of the greater flexibility of the coordination network. The increase in $\Delta \theta$ at low density, the typical measurement of amorphous order as inferred from the TO peak width in a Raman spectrum,\cite{Beeman} is also consistent with Jacks and Molina-Ruiz {\it et al.},\cite{Jacks,Molina} although we see a larger increase, perhaps due to finite-size effects of our supercell or limitations in the experimental extraction of $\Delta \theta$ and density in the films.
We find that $\Delta \theta$ increases at high densities also. The average bond angle decreases away from the relaxed density too, more dramatically for $a$-Si, which we will interpret in terms of effects near voids. The energies in CHASSM and DFT show increases away from the relaxed density, of course, but also a clear bump at the critical density for $a$-Si; no obvious feature occurs for $a$-Si:H.
A constant trend of $\langle r \rangle$ at low densities is consistent with the stabilized renormalized density in Fig. \ref{fig:voids}. These plots combined imply that Si-Si bonds have stopped stretching and begin to relax as a result of cavitation. Flattening of this atomic network density at low global densities is consistent with Rutherford backscattering spectroscopy and atomic force microscopy data.\cite{Jacks,Molina}

\begin{figure}
	\includegraphics[width=200px]{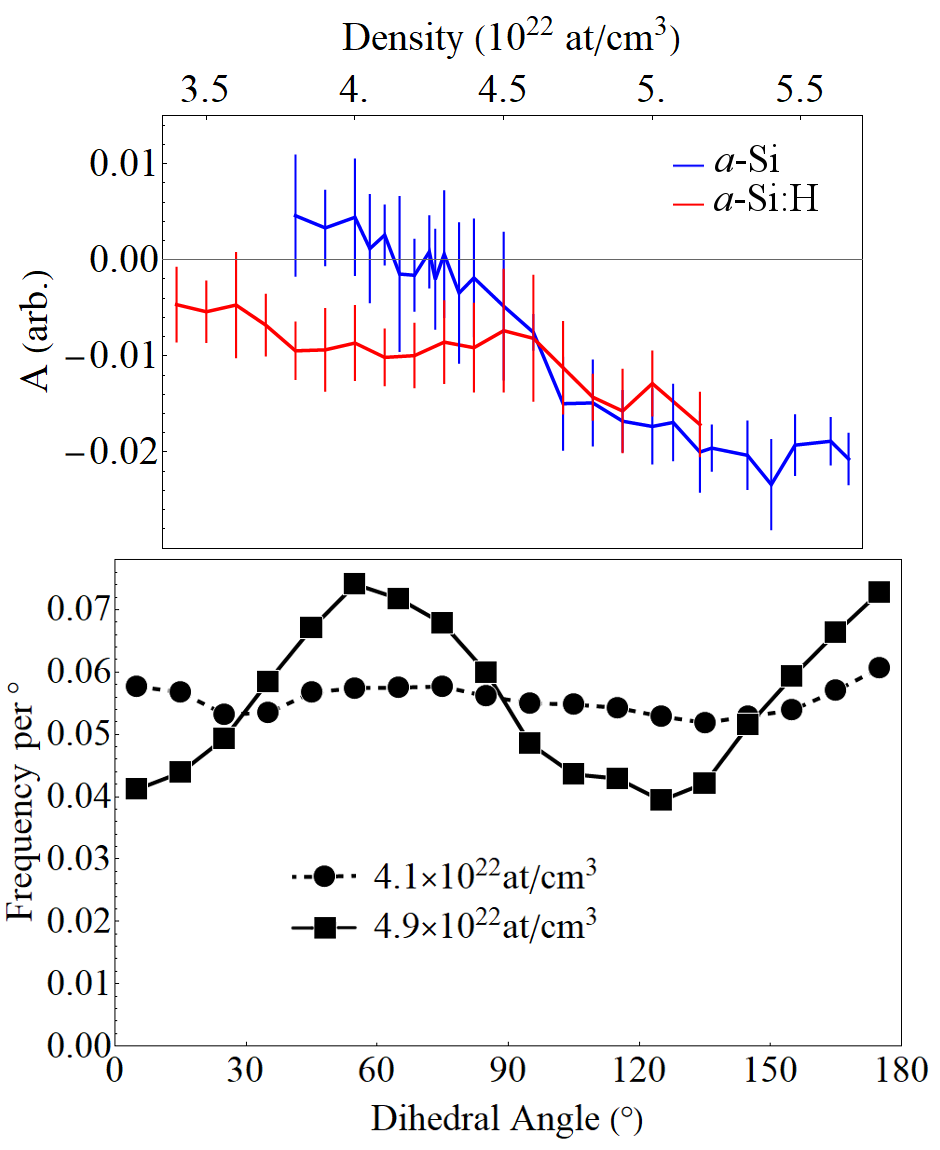}
	\caption{(top) The relationship between density and the dihedral oscillation amplitude, $A$. $A$ measures dihedral order, increasing as the density increases. (bottom) Dihedral distributions for a pair of low-density ($4.1\times$10$^{22}$ at/cm$^3$) and relaxed ($4.9\times$10$^{22}$ at/cm$^3$) $a$-Si structures. Dihedral order vanishes at the lowest densities.}
	\label{fig:Dih}
\end{figure}

Dihedral distributions show an unexpected density-induced variation. It is often considered that there is a uniform distribution of dihedrals in $a$-Si, inferred from the third nearest-neighbor peak in $g(r)$ as measured by X-ray diffraction.\cite{Schulke} However, our results show instead sinusoidal variation, with distinct peaks at $60^\circ$ and $180^\circ$, similar to what has been found in other computational studies\cite{Pedersen,Park,Holmstroem} and suggested by X-ray diffraction of $a$-Si.\cite{Laaziri}. For comparison, $c$-Si has $2/3$ of the dihedrals as $60^\circ$ and $1/3$ as $180^\circ$. To describe the density dependence, we restrict ourselves to Si atoms only and fit the dihedral distributions to the form $A \cos(2\pi \phi/120^\circ)+D$, where $\phi$ is the dihedral angle. $D$ is found to be density-invariant, but $A$, which we term the dihedral oscillation amplitude, is a measure of the dihedral order. Increasing the density increases the magnitude of $A$ (Fig. \ref{fig:Dih}), indicating a stronger medium range order at high densities. Lowest density structures show a flattening such that $A\rightarrow0$. In $a$-Si below $4.3\times10^{22}\ $at/cm$^3$, the relationship reverses and angles at $0^\circ$ and $120^\circ$ are more likely to be found than $60^\circ$ or $180^\circ$. Curiously, the lowest density $a$-Si structures with strong $0^\circ$ peaks are found to contain hexagonal bilayer sheets (like a graphene bilayer with AA stacking). We presume that these structures are unphysical artefacts of the Keating potential, and indeed the change of structure with DFT relaxation is increasingly large around these densities. Hexagonal bilayer sheets are compatible with large free surfaces while tetrahedrally coordinated structures necessarily suffer large deformations to their bond angles near a void. In $a$-Si:H, $A$ does not go above $0$ and we have not found evidence of hexagonal sheet structures.

\begin{figure}
	\includegraphics[width=380px]{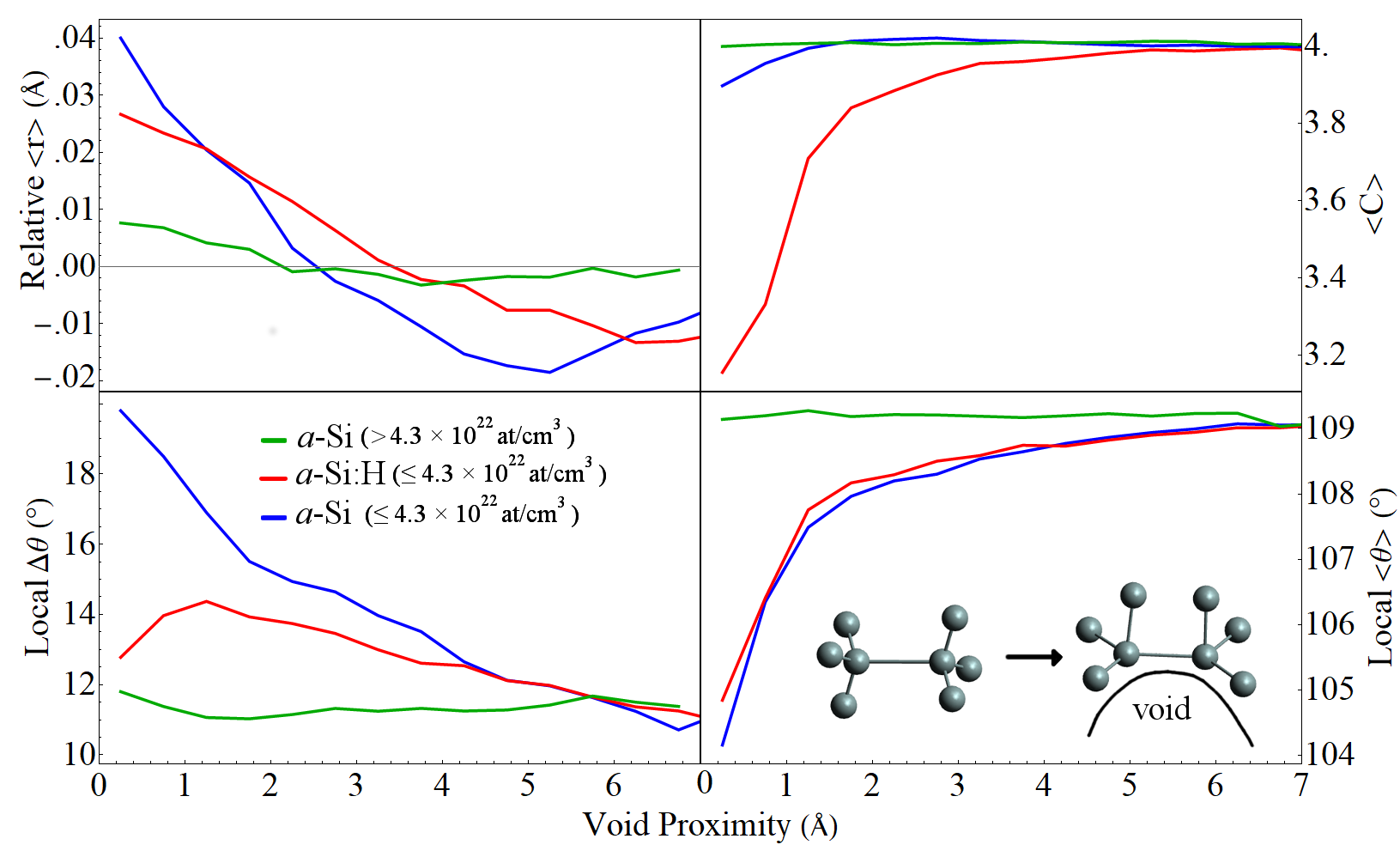}
	\caption{Locally resolved structure of low density structures as a function of void proximity: average bond length $\langle r \rangle$ with respect to global average bond length for the given density; average Si-Si coordination number $\langle C \rangle$; average bond angle deviation $\langle \Delta \theta \rangle$; and average bond angle $\langle \theta \rangle$. $a$-Si and $a$-Si:H lines are averaged over all structures with densities $4.3\times10^{22}\ $at/cm$^3$ and below. Higher-density structures (green) are plotted for comparison -- since the largest voids in these structures are not distinguishable from interstitials, there is little correlation between structure and void proximity. The results are consistent with a local rearrangement of bonds to accommodate a void as shown in the bottom-right sketch.}
	\label{fig:local}
\end{figure}

In low-density structures with large voids, structural deformations are associated with void proximity, $r_v$. To isolate local structural parameters, we group atoms based on their $r_v$ and collect bond lengths and angles associated with those atoms. $\Delta \theta$, $\langle \theta \rangle$, $\langle r \rangle$, and the average coordination number $\langle C \rangle$ are now computed on those sub-populations. Accurate description is limited by half the cell size minus the void diameter, to a distance of about 7 \AA\ away from a void surface. In a given low-density structure, the increased bond angle deviation resides entirely around the surface of voids  as shown in Fig. \ref{fig:local}. Carlson's model\cite{Carlson} of void surfaces resembling the $c$-Si (100) reconstruction is commonly invoked in the literature, \cite{Fehr} with respect to light-induced degradation. However, we do not see any resemblances in bond lengths and angles between that model and our the void surfaces. Results of 1-2\% increase in bond lengths and increasing local $\Delta \theta\ $are consistent with atomic removal methods.\cite{Kim} We have pictured a structural motif consistent with these structural changes in the lower right of Fig. \ref{fig:local}. Away from voids, $\Delta \theta$ returns to relaxed-like values of $10^\circ$. These results show conclusively that the structural changes below the $4.3\times10^{22}\ $at/cm$^3$ critical density are driven by voids.

Our low-density $a$-Si ensembles include structures that contain voids, despite maintaining perfect 4-coordination and having no Si-H bonds, at the cost of somewhat larger bond angle deviation. Such structures are not obtained by atomic removal methods, since 3-coordinated atoms are generated by design and there is a limited ability for void surfaces to reconstruct.\cite{Paudel} The existence of reasonable 4-coordinated void structures without Si-H bonds is significant as they are not detected in experiments such as IR studies or H effusion, requiring Si-H bonds, or electron spin resonance, requiring dangling bonds at the void surface. A range of degrees of H-passivation are likely to exist in voids.

Fig. \ref{fig:gr} and the $\langle C \rangle$ plot in Fig. \ref{fig:local} provide evidence of H clustering in $a$-Si:H. H atoms are highly concentrated within voids, especially at low densities. This is significant since we did not explicitly place H atoms at the void surfaces, as in previous work,\cite{Chakraborty,Pedersen,Nakhmanson,Biswas_2007,Paudel} but the H atoms naturally ended up there from the Monte Carlo process and annealing. This result is consistent with previous studies.\cite{Chakraborty,Wright,Ouwens}

Finally, ring analysis (calculated using King's method\cite{King} with the open-source code R.I.N.G.S.\cite{RINGS}) are consistent with previous works.\cite{Deringer} There is little density-dependence in the ring statistics of $a$-Si:H. An increase in $a$-Si six-membered rings at the lowest densities is present---consistent with the observed hexagonal sheets.

\section{Conclusion}
Using a WWW-based Monte Carlo method with different fixed densities, followed by DFT relaxation, we are able to generate realistic $a$-Si and $a$-Si:H structures with voids that arise as equilibrium structures at densities below a critical density. These structures can be used to study effects of voids on degradation, hole mobility, two-level system phenomena, thermal transport, or elastic properties. The method is simple and scales well with system size,\cite{Barkema} and is efficient by focusing specifically on the voids and their immediate neighborhoods. Our approach requires no atomic addition or removal, nor any {\it a priori} idea of the targeted structures.  We verified the validity of the Keating potential description across a range of densities around the relaxed one, except for the most extreme densities studied.

We find in $a$-Si:H that H atoms tend to be concentrated near voids. By contrast, our method is unique for obtaining $a$-Si structures with voids that have near-perfect coordination without any H passivation. Similar fully coordinated $a$-Si void structures may exist and be overlooked in experiments that assume Si-H bonds or dangling bonds at voids. Nakhmanson and Drabold\cite{Nakhmanson} found low-energy vibrational modes localized near void surfaces. These may be a product of low coordination near voids produced by atomic removal and may be worth revisiting with fully 4-coordinated pure $a$-Si void structures. These phonon modes may have strong implications on two-level systems.\cite{Molina_2018}

Our structural analysis of $a$-Si and $a$-Si:H at low density shows a collection of changes connected to void formation: an increase in bond lengths, and then a decrease; a decrease in average bond angle; an increase in bond angle deviation (less tetrahedral bond angles); and a decrease in medium-range dihedral order. Increasing density above the equilibrium value also decreases bond angles and increases bond angle deviation. $a$-Si:H responds more smoothly to strain than $a$-Si due to a less constrained network. The increase in negative pressure and then reduction below the critical density indicates the bonding network undergoes a bubble-like cavitation process---the formation of large voids to relieve internal stresses. By resolving structures at an atomic level, we conclude that the structural changes at low density reside near void surfaces. The locality of this structural disorder may be related to two-level systems.\cite{Molina_2018} Our approach for void generation opens the way to realistic studies of void effects on the photovoltaic, electronic, thermal, and mechanical properties of $a$-Si and $a$-Si:H. We believe this approach to generating voids can be useful also for studies of other amorphous materials, and help elucidate fundamental questions that remain about physics in the amorphous state.



\section{Acknowledgments}
We acknowledge useful discussions with Gergely Zim\'{a}nyi, Frances Hellman, Manel Molina-Ruiz, and Hilary Jacks. This work was supported by UC Merced start-up funds and by the Merced nAnomaterials Center for Energy and Sensing (MACES), a NASA-funded research and education center, under award NNX15AQ01. This work used computational resources from the Multi-Environment Computer for Exploration and Discovery (MERCED) cluster at UC Merced, funded by National Science Foundation Grant No. ACI-1429783, and the National Energy Research Scientific Computing Center (NERSC), a U.S. Department of Energy Office of Science User Facility operated under Contract No. DE-AC02-05CH11231.

\end{document}